\documentclass[submission, Proceedings]{SciPost}

\hypersetup{
    colorlinks,
    linkcolor={red!50!black},
    citecolor={blue!50!black},
    urlcolor={blue!80!black}
}

\usepackage[bitstream-charter]{mathdesign}
\DeclareSymbolFont{usualmathcal}{OMS}{cmsy}{m}{n}
\DeclareSymbolFontAlphabet{\mathcal}{usualmathcal}

\usepackage{caption}
\usepackage{subcaption}

\usepackage{physics}

\newcommand{\rt}{{\mathbf{r}}}

\newcommand{\bt}{{\mathbf{b}}}

\newcommand{\xpom}{{x_\mathbb{P}}}

\newcommand{\fig}{Fig.~}

\newcommand{\der}{\mathrm{d}}

\newcommand{\Deltat}{\mathbf{\Delta}}

\newcommand{\jpsi}{$\mathrm{J}/\psi$ }

\begin{document}

\begin{center}{\Large \textbf{
Higher-order corrections to exclusive heavy vector meson production\\
}}\end{center}

% TODO: write the author list here. Use initials + surname format.
% Separate subsequent authors by a comma, omit comma at the end of the list.
% Mark the corresponding author with a superscript *.
\begin{center}
T. Lappi\textsuperscript{1,2},
H. Mäntysaari\textsuperscript{1,2} and
J. Penttala\textsuperscript{1,2$\star$}
\end{center}

\begin{center}
{\bf 1} Department of Physics, University of Jyväskylä, P.O. Box 35, 40014 University of Jyväskylä, Finland
\\
{\bf 2} Helsinki Institute of Physics, P.O. Box 64, 00014 University of Helsinki, Finland
\\
* jani.j.penttala@student.jyu.fi
\end{center}

\begin{center}
\today
\end{center}

% For convenience during refereeing (optional),
% you can turn on line numbers by uncommenting the next line:
%\linenumbers

\definecolor{palegray}{gray}{0.95}
\begin{center}
\colorbox{palegray}{
  \begin{tabular}{rr}
  \begin{minipage}{0.1\textwidth}
    \includegraphics[width=22mm]{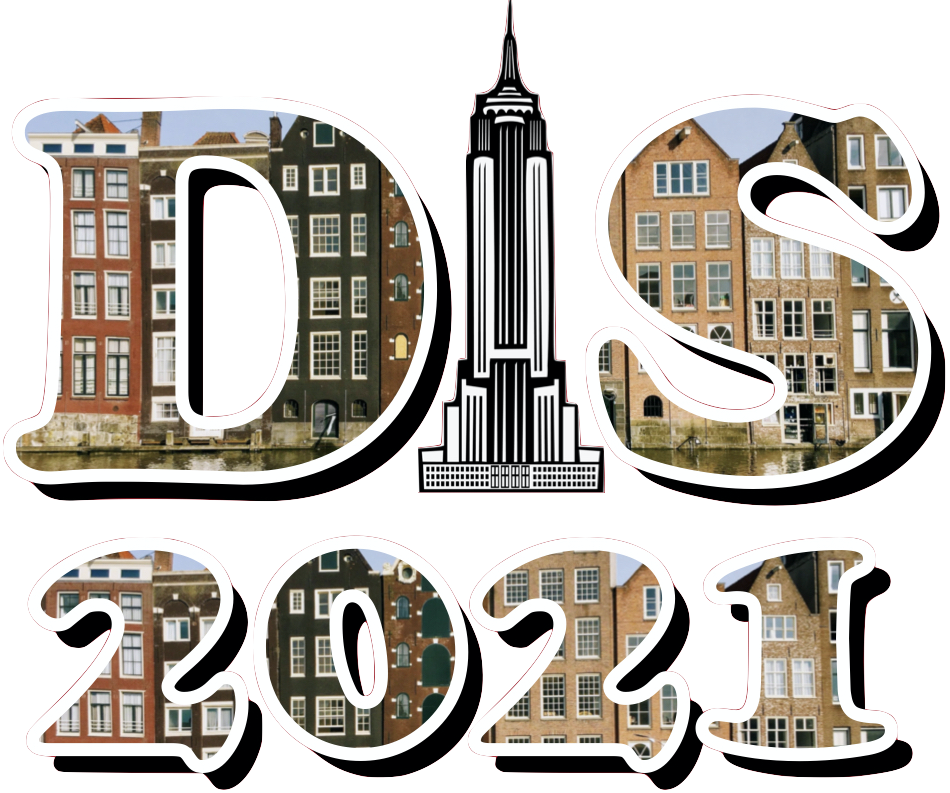}
  \end{minipage}
  &
  \begin{minipage}{0.75\textwidth}
    \begin{center}
    {\it Proceedings for the XXVIII International Workshop\\ on Deep-Inelastic Scattering and
Related Subjects,}\\
    {\it Stony Brook University, New York, USA, 12-16 April 2021} \\
    \doi{10.21468/SciPostPhysProc.?}\\
    \end{center}
  \end{minipage}
\end{tabular}
}
\end{center}

\section*{Abstract}
{\bf

We present results for higher-order corrections to exclusive \jpsi production. This includes the first relativistic correction of order $v^2$ in quark velocity, and next-to-leading order corrections in $\alpha_s$ for longitudinally polarized production. The relativistic corrections are found to be important for a good description of the HERA data, especially at small values of the photon virtuality. The next-to-leading order results for longitudinal production are evaluated numerically. We also demonstrate how the vector meson production provides complementary information to the structure functions for extracting the initial condition for the small-$x$ evolution of the dipole-proton scattering amplitude.
}

% \vspace{10pt}
% \noindent\rule{\textwidth}{1pt}
% \tableofcontents\thispagestyle{fancy}
% \noindent\rule{\textwidth}{1pt}
% \vspace{10pt}

\section{Introduction}
\label{sec:intro}
Exclusive vector meson production is a powerful tool for determining the small-$x$ gluon structure of the proton. There is already plenty of data for this process from HERA, and in the future this will be supplemented by measurements at the LHC and the future Electron-Ion Collider (EIC)~\cite{AbdulKhalek:2021gbh}. Thus there is a need for a precise theoretical description of the process. Currently, there is a major source of model uncertainty coming from the vector meson wave function which is often taken from phenomenological models or assumed to be fully nonrelativistic. In this Contribution, we will consider the first relativistic corrections to the heavy vector meson wave function, which will quantify model uncertainties of the wave function. We will also present results of the next-to-leading order calculation for exclusive production of longitudinally polarized heavy vector mesons, which is important for developing a precise theoretical description of the process.

\section{Relativistic corrections of order $v^2$}

At high energies exclusive vector meson production can be conveniently described in the dipole picture where an energetic virtual photon scatters off the target proton or  nucleus and forms the vector meson. This allows us to describe the process in a factorized form where the scattering amplitude reads~\cite{Kowalski:2006hc}
\begin{equation}
\label{eq:vm_amp}
    \mathcal{A}^\lambda = 2i\int \der^2 \bt \der^2 \rt \frac{\der z}{4\pi} e^{-i \left(\bt + \left(\frac{1}{2}-z\right)\rt\right) \cdot \mathbf{\Delta}} 
    \Psi_\gamma^{\lambda *}(\rt,Q^2,z) \Psi_V(\rt,z) N(\rt,\bt,\xpom)
\end{equation}
for a photon with polarization $\lambda$. Here the process is described in a mixed coordinate space where $\rt$ is the transverse size of the quark-antiquark dipole and $z$ is the longitudinal momentum fraction carried by the quark. The necessary parts for the calculation are the virtual photon wave function $\Psi_\gamma^{\lambda}$, the meson wave function $\Psi_V$, and the dipole amplitude $N$. The photon wave function can be calculated directly by using light-front perturbation theory~\cite{Dosch:1996ss}. For the dipole amplitude we use here the IPSat parametrization~\cite{Kowalski:2003hm} with the fit from Ref.~\cite{Mantysaari:2018nng}. This particular choice for the dipole amplitude is dependent on the impact parameter $\bt$ which allows us to study the dependence of the scattering amplitude on the momentum transfer $t \approx -\Deltat^2$.

The final part needed for the scattering amplitude, the meson light front wave function, is nonperturbative and therefore introduces an additional model dependence. 
%The meson wave function is therefore usually taken from phenomenological models, such as in Refs.~\cite{Kowalski:2006hc,Li:2020ntd}, or assumed to be completely nonrelativistic. 
In Ref.~\cite{Lappi:2020ufv}, we propose a new approach for the meson wave function based on Non-Relativistic QCD (NRQCD), which includes relativistic corrections of order $v^2$ in the quark's velocity $v$. In this approach, we start from the rest frame of the meson where the value of the wave function and its derivatives can be written in terms of the NRQCD long-distance matrix elements. This procedure allows us to include relativistic corrections of order $v^2$ to the meson's rest frame wave function in terms of two nonperturbative constants, related to the NRQCD matrix elements, that have been determined for \jpsi from the decay constants of charmonia in Ref.~\cite{Bodwin:2007fz}. The meson rest frame wave function is then related to the light front wave function by the so-called Melosh rotation~\cite{Melosh:1974cu} which acts as a transformation between two different spinor bases.

\begin{figure}
  \centering
  \begin{subfigure}{0.49\textwidth}
      \centering
      \includegraphics[width=\textwidth]{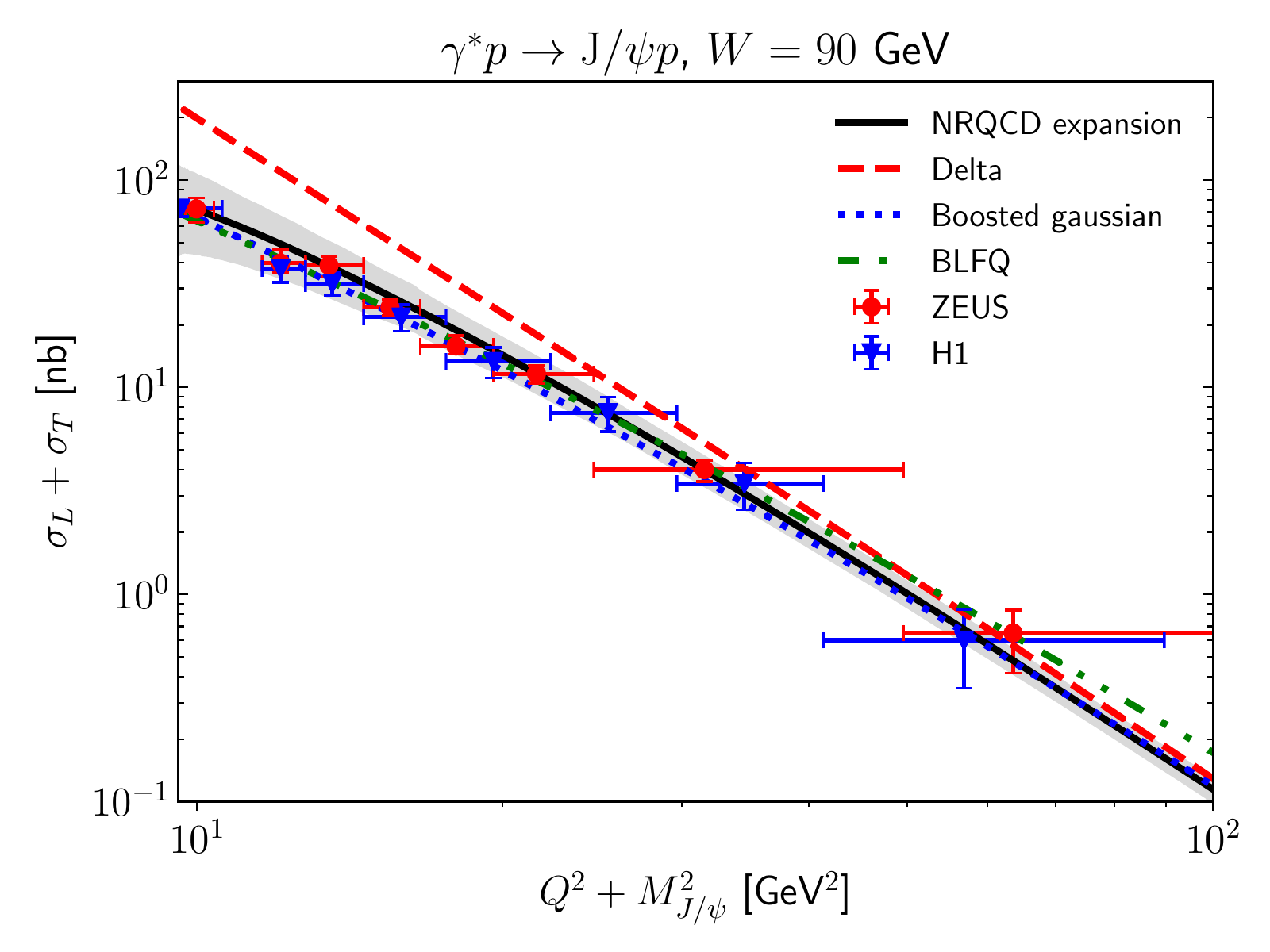}
  \end{subfigure}
  \caption{Total exclusive \jpsi production cross section as a function of the photon virtuality~\cite{Lappi:2020ufv}. 
  }
  \label{fig:rel_corrections}
\end{figure}

To quantify the effects of the relativistic corrections, we show the total \jpsi production cross section as a function of the photon virtuality $Q^2$ in \fig\ref{fig:rel_corrections}. Here we compare results from four different meson wave functions: the fully nonrelativistic (\emph{Delta}), one with relativistic corrections of order $v^2$ (\emph{NRQCD expansion}), and two phenomenological wave functions from Refs.~\cite{Kowalski:2006hc,Li:2020ntd}. These are compared to the HERA data from Refs.\cite{Aktas:2005xu,Chekanov:2004mw}. We see that the relativistic corrections have the greatest impact at small values of the photon virtuality, whereas they become negligible at large $Q^2$. The relativistic corrections are also crucial for a good description of the HERA data as the fully nonrelativistic wave function results in an overestimation of the total cross section at small $Q^2$.

\section{Next-to-leading order corrections at the nonrelativistic limit}
Of comparative importance are the next-to-leading order corrections to the vector meson production. The significance of the relativistic and $\alpha_s$ corrections can be estimated by NRQCD which tells us that numerically $v^2 \sim \alpha_s$ in the case of \jpsi \cite{Bodwin:2007fz}. This means that we can do an expansion in both the velocity and $\alpha_s$, which indicates that the first NLO correction should be calculated at the nonrelativistic limit~\cite{Escobedo:2019bxn}. The calculation of the NLO corrections requires that we know the virtual photon and meson wave functions at the corresponding accuracy. In recent years, there has been major progress in calculating these: the NLO corrections to the vector meson wave function \cite{Escobedo:2019bxn} and to the longitudinal photon wave function with massive quarks \cite{Beuf:2021qqa} have recently become available.

\begin{figure}
  \centering
  \begin{subfigure}{0.49\textwidth}
      \centering
      \includegraphics[width=\textwidth,keepaspectratio]{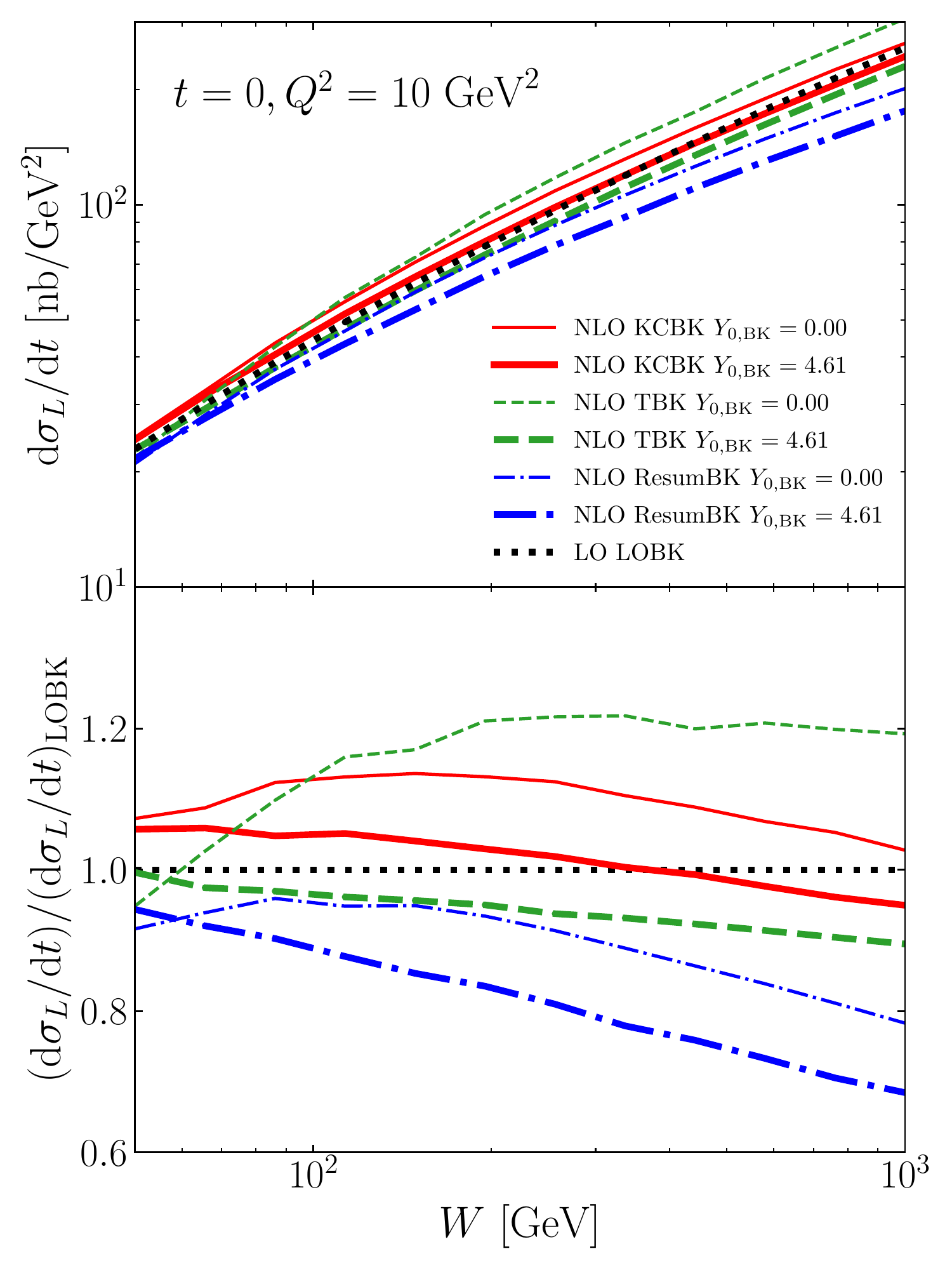}
  \end{subfigure}
  \hfill
  \begin{subfigure}{0.49\textwidth}
      \centering
      \includegraphics[width=\textwidth,keepaspectratio]{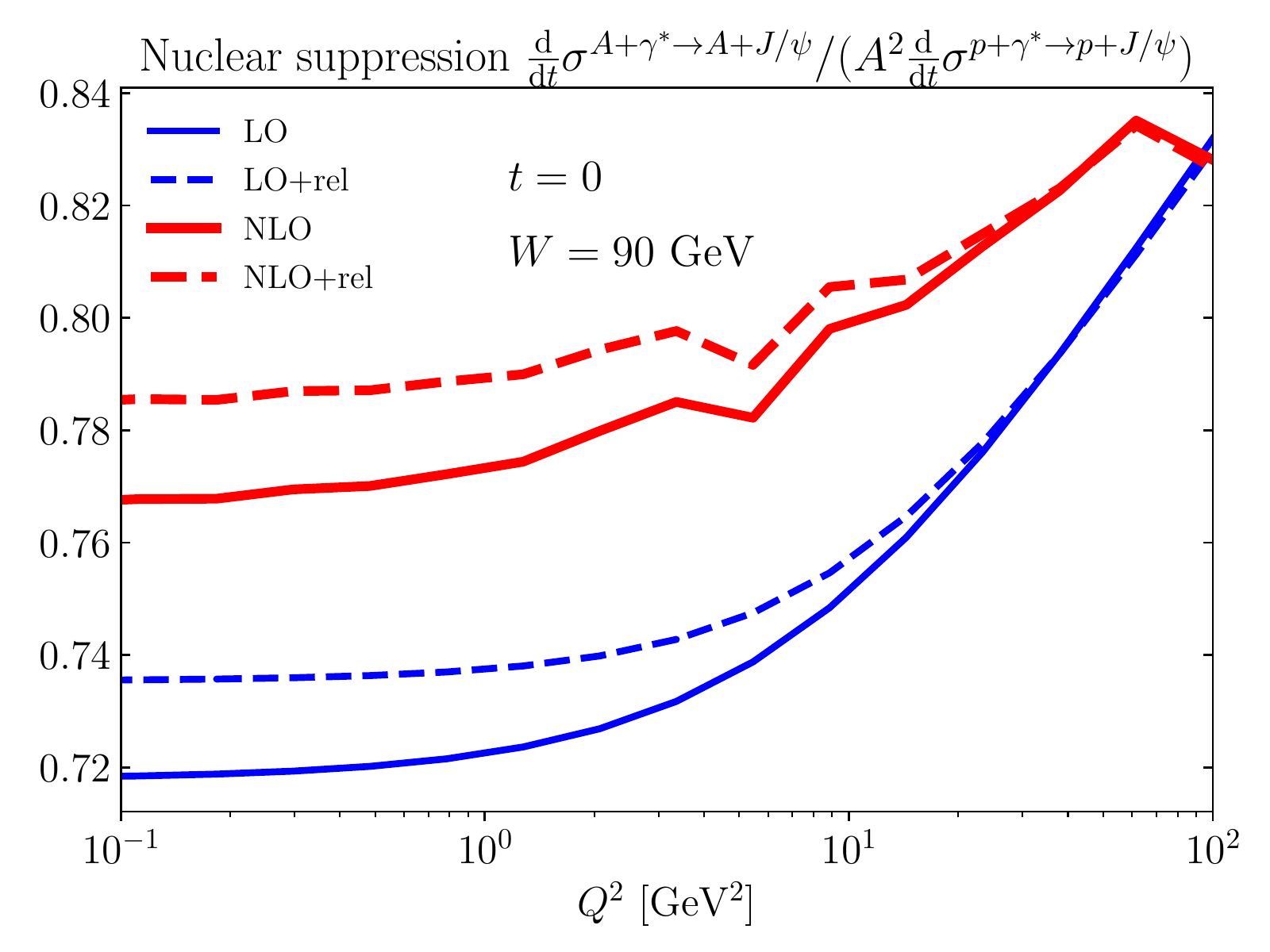}
  \end{subfigure}
  \caption{
  Left: Diffractive cross section for longitudinal \jpsi production as a function of the center-of-mass energy~\cite{Mantysaari:2021ryb}. \emph{LO BK} corresponds to a leading-order result with a leading-order dipole amplitude fit. The other curves show the NLO results with different NLO dipole amplitude fits. In the lower plot, the curves have been normalized by the leading-order result.  \\
  Right: Nuclear suppression for longitudinal \jpsi production. We compare the LO and NLO results at the nonrelativistic limit and with the relativistic corrections of order $v^2$. }
  \label{fig:nlo_corrections}
\end{figure}

In Ref.~\cite{Mantysaari:2021ryb}, we combined these NLO wave functions to calculate exclusive \jpsi production at NLO in the longitudinal polarization case. This calculation demonstrates the cancellation of all UV and IR divergences present in the wave functions, and shows explicitly the appearance of the Balitsky-Kovchegov (BK) equation for the energy dependence of the dipole amplitude. In \fig\ref{fig:nlo_corrections}, we show the numerical results for the NLO calculation of \jpsi production. In the left plot, the results for diffractive cross sections from the LO and NLO calculations are shown. The dipole amplitudes used in these calculations are taken from~\cite{Lappi:2013zma} in the LO case and~\cite{Beuf:2020dxl} in the NLO case, and in both cases they have been fitted to the HERA structure function data. For the NLO case, we have calculated the results using various different parametrizations for the dipole amplitude satisfying modified versions of the BK equation, abbreviated here as \emph{KCBK}, \emph{TBK}, and \emph{ResumBK}. This figure shows that there is more variation between the different NLO results than between the LO and NLO results. This tells us that vector meson production provides complementary information to structure function analyses, since all of these dipole amplitudes have been successfully fitted to the structure function data.

In the right plot, we show a ratio of the cross sections for the nuclear and proton targets. This ratio measures the nuclear suppression as without nonlinear effects the ratio would be identically one. We have compared the results from the LO and NLO cases, with and without relativistic corrections of order $v^2$. The effects of the NLO and relativistic corrections seem to be small, although they both reduce the nuclear suppression slightly. In Ref.~\cite{Lappi:2020ufv}, the effects of the relativistic corrections to nuclear suppression were found to be larger. The differences could be due to the fact that here we are only considering the longitudinal diffractive cross section at $t=0$ as opposed to the total cross section, and a different dipole amplitude was used in the two calculations.

\section{Conclusion}

We have shown how the first relativistic correction and the NLO correction in the longitudinal polarization case affect exclusive production of \jpsi. The relativistic corrections are important at small $Q^2$ for a good agreement with the HERA data. The NLO corrections affect the cross section only modestly, and there is more variation between the results using different NLO fitted dipole amplitudes than between the LO and NLO results.

In the future, we are going to include next-to-leading order calculations for transversely polarized virtual photons. This will enable us to make comparisons to the vector meson production data from HERA~\cite{Alexa:2013xxa,Chekanov:2002xi,Chekanov:2004mw} and the LHC~\cite{Aaij:2014iea,Acharya:2018jua}, while also making precise predictions for the future EIC.

% \section*{Acknowledgements}

% \paragraph{Author contributions}
% This is optional. If desired, contributions should be succinctly described in a single short paragraph, using author initials.

% TODO: include funding information
\paragraph{Funding information}
This  work  was  supported  by  the  Academy  of  Finland, projects 314764 (H.M), 321840 (T.L and J.P) and 314162 (J.P),  and by the EU Horizon 2020 research and innovation programme, STRONG-2020 project (Grant Agreement No. 824093). T.L is supported by the European Research Council (ERC) under the European Union’s Horizon 2020 research and innovation programme (grant agreement No. ERC-2015-CoG-681707).  The content of this article does not reflect the official opinion of the European Union and responsibility  for  the  information  and  views  expressed therein lies entirely with the authors.

% TODO:
% Provide your bibliography here. You have two options:

% FIRST OPTION - write your entries here directly, following the example below, including Author(s), Title, Journal Ref. with year in parentheses at the end, followed by the DOI number.
%\begin{thebibliography}{99}
%\bibitem{1931_Bethe_ZP_71} H. A. Bethe, {\it Zur Theorie der Metalle. i. Eigenwerte und Eigenfunktionen der linearen Atomkette}, Zeit. f{\"u}r Phys. {\bf 71}, 205 (1931), \doi{10.1007\%2FBF01341708}.
%\bibitem{arXiv:1108.2700} P. Ginsparg, {\it It was twenty years ago today... }, \url{http://arxiv.org/abs/1108.2700}.
%\end{thebibliography}

% \bibliographystyle{SciPost_bibstyle} % Include this style file here only if you are not using our template
\bibliography{refs.bib}

\nolinenumbers

\end{document}